\documentclass[12pt]{article}

\usepackage{epsfig}
\usepackage{array}
\usepackage{multirow}
\usepackage{longtable}
\usepackage{supertabular}
\usepackage{amssymb}
\usepackage{times}

\begin{document}

\title{Modelling droplets on superhydrophobic surfaces:
       equilibrium states and transitions}

\author{A.~Dupuis and J.M.~Yeomans \\
   \footnotesize{The Rudolf Peierls Centre for Theoretical Physics, University of Oxford,}\\
 \footnotesize{1 Keble Road, Oxford OX1 3NP, UK.}}

\maketitle

\begin{abstract}
We present a lattice Boltzmann solution of the equations of motion
describing the spreading of droplets on topologically patterned
substrates. We apply it to model superhydrophobic behaviour on
surfaces covered by an array of micron-scale posts. We find that the
patterning results in a substantial increase in contact angle, from
$110^o$ to $156^o$. The dynamics of the transition from drops
suspended on top of the posts to drops collapsed in the grooves is
described.
\end{abstract}

\newcommand{\pos}{\ensuremath{\mathbf{r}}}
\newcommand{\dt}{\ensuremath{\Delta t}}
\newcommand{\dr}{\ensuremath{\Delta \pos}}
\newcommand{\vi}{\ensuremath{\mathbf{v}_i}}
\newcommand{\vm}{\ensuremath{v}}
\renewcommand{\u}{\ensuremath{\mathbf{u}}}
\newcommand{\dab}{\ensuremath{\delta_{\alpha\beta}}}
\newcommand{\eq}[1]{equation (\ref{#1})}
\newcommand{\Eq}[1]{Equation (\ref{#1})}
\newcommand{\fig}[1]{fig. \ref{#1}}
\newcommand{\Fig}[1]{Fig. \ref{#1}}

% ============================================================
\section{Introduction}
\label{sec:intro}

From  microfluidic technology to detergent design and ink-jet printing
it is vital to understand the way in which drops move across
surfaces. The dynamics of the drops will be affected by any chemical or
topological heterogeneities on the surface. Until recently such
disorder was usually regarded as undesirable. However with the advent
of microfabrication techniques it has become possible to control the
chemical or topographical patterning of a substrate on micron length
scales, leading to the possibility of exploring new physics and to
novel applications.

%A droplet in contact with a substrate will try to spread to an
%equilibrium shape determined by Young's law which describes the
%balance of surface tensions. There are many parameters which affect
%this process. For example surface disorder in the form of chemical or
%topological heterogeneities can pin a droplet or change its final
%shape. This has usually been viewed as a nuisance in experiments and
%applications. However with the advent of microfabrication techniques
%it is becoming possible to harness controlled surface topologies to
%explore new physical phenomena.

A beautiful example of this, inspired by the leaves of the lotus
plant, are superhydrophobic substrates. These are surfaces which are
covered with posts on length scales of order microns. As a result of
the topological patterning they are strongly repellent to water
droplets which show contact angles up to
$160^{o}$~\cite{bico:99,oner:00,he:03}. This should be compared to
more traditional ways of increasing the contact angle, surface
coatings and chemical modifications of the substrate, where it is
difficult to achieve an angle of more than $120^{o}$.
Superhydrophobic substrates have many potential applications, for
example as materials for raincoats or windscreens. The evolutionary
advantage to the lotus appears to be the easy run-off which helps to
clean the leaves of the plant.

Droplets on a superhydrophobic surface can be in two states, suspended
where the drop sits on top of the posts with pockets of air beneath
it, or collapsed where it wets the grooves between the posts.  Several
authors have calculated the free energies of the suspended and
collapsed states using approximations based on the Cassie-Baxter and
Wenzel laws respectively~\cite{patankar:04,ishino:04}. They have shown
that both states can be thermodynamically stable with the phase
boundary between them depending on the surface tension and substrate
geometry. It has also been argued that the suspended drop is often
observed as a metastable state as it has to cross a free energy
barrier to fill the grooves. However the kinetics of the transition to
the collapsed phase is not understood: it is not accessible to the
equilibrium theories and it is hard to probe experimentally.

Therefore in this paper we present a numerical solution to equations
which are able to describe both the static and dynamic behaviour of
droplets on topologically patterned substrates. The droplet dynamics
is described by the Navier-Stokes equations for a liquid-gas
system. Its equilibrium behaviour corresponds to a chosen free energy
functional so that appropriate thermodynamic information, such as the
surface tension and the contact angle, are included in the model. We
solve the equations of motion using a lattice Boltzmann
algorithm. This approach has a natural length scale, for fluids such
as water, of order microns where much of the exciting new physics is
expected to appear. The method has already shown its capability in
dealing with spreading on surfaces with chemical
patterning~\cite{leopoldes:03}.

In section 2 we summarise the algorithm and, particularly, describe
the new thermodynamic and velocity boundary conditions needed to treat
surfaces with topological patterning. In section 3 we present results
for a substrate patterned by an array of posts and show that the
patterning leads to a substantial increase in contact angle in
agreement with experiments. We then explore the kinetics of the
transition between the suspended and collapsed droplet states.
Finally we discuss directions for future work using this approach.

% ============================================================
\section{The mesoscopic model}
\label{sec:model}

\subsection{Equilibrium free energy}

We consider a liquid-gas system of density $n(\pos)$ and volume
$V$. The surface of the substrate is denoted by $S$. The equilibrium
properties of the drop are described by the free energy
\begin{equation}
\Psi = \int_V \left( \psi_b(n) + \frac{\kappa}{2} \left( \partial_\alpha n
  \right)^2 \right) dV + \int_S \psi_c(n) \; dS.
\label{eq:freeE}
\end{equation}
where Einstein notation is understood for the Cartesian label $\alpha$
(i.e.  $v_{i\alpha}u_\alpha=\sum_\alpha v_{i\alpha}u_\alpha$) and
where $\psi_b(n)$ is the free energy in the bulk. We conveniently
choose the double well form~\cite{briant:04}
\begin{equation}
\psi_b(n)=p_c \left( \nu_n+1 \right)^2 (\nu_n^2-2\nu_n+3-2\beta\tau_w)
\label{eq:freeEbulk}
\end{equation}
where $\nu_n=(n-n_c)/n_c$, $\tau_w=(T_c-T)/T_c$ and $p_c=1/8$,
$n_c=7/2$ and $T_c=4/7$ are the critical pressure, density and
temperature respectively and $\beta$ is a constant typically equal to
$0.1$.

The derivative term in \eq{eq:freeE} models the free energy associated
with density gradients at an interface. $\kappa$ is related to the
surface tension $\gamma$ by~\cite{briant:04}
\begin{equation}
\gamma=\frac{4}{3} \sqrt{2 \kappa p_c} (\beta\tau_w)^{3/2}n_c.
\end{equation}

$\psi_c(n_s)=-\phi_1 n_s$ is the Cahn surface free
energy~\cite{cahn:77} which controls the wetting properties of the
fluid. In particular $\phi_1$ can be used to tune the contact angle.
 
%In this framework the surface tension $\gamma$ is given by
%\begin{equation}
%\gamma=\frac{4}{3} \sqrt{2 \kappa p_c} (\beta\tau_w)^{3/2}n_c.
%\label{eq:st}
%\end{equation}

\subsection{Navier-Stokes equations}

The dynamics of the droplet is described by the Navier-Stokes equations
for a non-ideal gas
\begin{eqnarray}
\partial_t (n u_\alpha) + \partial_\beta (nu_\alpha u_\beta) & = & 
  - \partial_\beta P_{\alpha\beta} 
  + \nu \partial_\beta \left[ n (\partial_\beta u_\alpha + 
                              \partial_\alpha u_\beta + 
                              \delta_{\alpha\beta} \partial_\gamma u_\gamma) 
                       \right] + nF_\alpha, \nonumber \\
\partial_t n + \partial_\alpha(n u_\alpha) & = & 0
\label{eq:ns} 
\end{eqnarray}
where $\mathbf{u}(\pos)$ is the fluid velocity, $\nu$ the kinematic
viscosity and $\mathbf{F}$ a gravitational field. The pressure tensor
$P_{\alpha\beta}$ is related to the free energy by~\cite{anderson:98}
\begin{equation}
P_{\alpha\beta} = \frac{\partial \mathcal{F}}{\partial (\partial_\alpha n)} (\partial_\beta n) - \mathcal{F} \delta_{\alpha\beta}
\end{equation}
where $\mathcal{F}=\psi_b - \mu_b n + \kappa(\partial_\alpha n)^2/2$
and $\mu_b=4p_c(1-\beta\tau_w)/n_c$ is the bulk chemical potential.

%The lattice Boltzmann algorithm solves the Navier-Stokes equations for
%this system. Because interfaces appear naturally within the model it
%is particularly well suited to looking at the behaviour of moving
%drops.

\subsection{The lattice Boltzmann algorithm}

\begin{figure}
\begin{center}
\epsfig{file=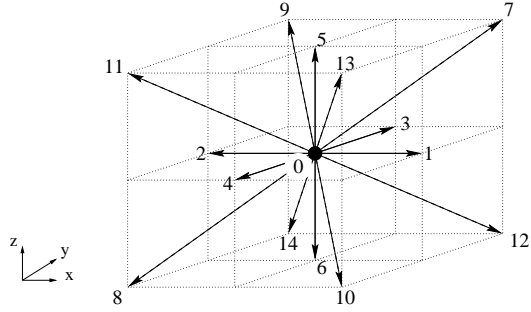,width=7cm}
\end{center}
\caption{Topology of a $D3Q15$ lattice. The directions $i$ are 
numbered and correspond to the velocity vectors $\vi$.}
\label{fig:d3q15}
\end{figure}

We solve the equations of motion (\ref{eq:ns}) by using a lattice
Boltzmann algorithm. This approach follows the evolution of partial
distribution functions $f_i$ on a regular, $d$-dimensional lattice
formed of sites $\pos$. The label $i$ denotes velocity directions and
runs between $0$ and $z$. $DdQz+1$ is a standard lattice topology
classification. The $D3Q15$ lattice we use here has the following
velocity vectors $\vi$: $(0,0,0)$, $(\pm 1,\pm 1,\pm 1)$, $(\pm
1,0,0)$, $(0,\pm 1, 0)$, $(0,0, \pm 1)$ in lattice units as shown in
\fig{fig:d3q15}.

The lattice Boltzmann dynamics are given by
\begin{equation}
f_i(\pos+ \dt \vi,t+\dt)=f_i(\pos,t)+\frac{1}{\tau}\left(f_i^{eq}(\pos,t)-f_i(\pos,t)\right) + n w_\sigma v_{i\alpha} F_\alpha
\label{eq:lbDynamics}
\end{equation}
where $\dt$ is the time step of the simulation, $\tau$ the relaxation
time. $\sigma$ labels velocities of different magnitude, $w_1=1/3$,
$w_2=1/24$. $f_i^{eq}$ is the equilibrium distribution function which
is a function of the density $n=\sum_{i=0}^z f_i$ and the fluid
velocity $\u$, defined through the relation
\begin{equation}
n\u=\sum_{i=0}^z f_i\vi.
\label{lb:velocity}
\end{equation}

\noindent
The relaxation time tunes the kinematic viscosity as
\begin{equation}
\nu=\frac{(\dr)^2}{\dt} \frac{1}{3} (\tau-\frac{1}{2})
\label{eq:visco}
\end{equation}
where $\dr$ is the lattice spacing.

It can be shown~\cite{swift:96} that equation~(\ref{eq:lbDynamics})
reproduces the Navier-Stokes equations of a non-ideal gas
(\ref{eq:ns}) if the local equilibrium functions are chosen as
\begin{eqnarray}
f_i^{eq}&=&A_\sigma+B_\sigma u_\alpha v_{i\alpha} + C_\sigma \u^2
         +D_\sigma u_\alpha u_\beta v_{i\alpha}v_{i\beta} 
         +G_{\sigma\alpha\beta} v_{i\alpha}v_{i\beta}, \quad i>0,
         \nonumber \\
f_0^{eq}&=& n - \sum_{i=1}^z f_i^{eq}.
\label{eq:lbEq}
\end{eqnarray}
\noindent
A possible choice of the coefficients is~\cite{dupuis:03c}
\begin{eqnarray}
A_\sigma & = & \frac{w_\sigma}{c^2}\left( p_0- 
               \frac{\kappa}{2} (\partial_\alpha n)^2               
               -\kappa n \partial_{\alpha\alpha} n 
               + \nu u_\alpha \partial_\alpha n \right), \nonumber \\
B_\sigma & = & \frac{w_\sigma n}{c^2}, \quad 
  C_\sigma = -\frac{w_\sigma n}{2 c^2}, \quad 
  D_\sigma = \frac{3 w_\sigma n}{2 c^4}, \nonumber \\
G_{1\gamma\gamma} & = & \frac{1}{2 c^4} \left( \kappa(\partial_\gamma n)^2 +2 
\nu u_\gamma \partial_\gamma n \right) , \quad 
  G_{2\gamma\gamma} = 0, \nonumber \\
G_{2\gamma\delta} & = & \frac{1}{16 c^4} \left( \kappa (\partial_\gamma n)
  (\partial_\delta n) + \nu (u_\gamma \partial_\delta n + u_\delta
 \partial_\gamma n) \right)
\label{lb:eqCoeff}
\end{eqnarray}
where $c=\dr/\dt$ and $p_0=n \partial_n \psi_b - \psi_b =
p_c(\nu_n+1)^2(3\nu_n^2-2\nu_n+1-2\beta\tau_w)$.

\subsection{Wetting boundary conditions}

The major challenge in dealing with patterned substrates is to handle
the boundary conditions correctly. We consider first wetting boundary
conditions which control the value of the density derivative and hence
the contact angle. For flat substrates a boundary condition can be
established by minimising the free energy
(\ref{eq:freeE})~\cite{cahn:77}
\begin{equation}
\hat{\mathbf{s}} \cdot \nabla n = - \frac{\phi_1}{\kappa}
\label{eq:cahn1}
\end{equation}
where $\hat{\mathbf{s}}$ is the unit vector normal to the substrate.
It is possible to obtain an expression relating $\phi_1$ to the
contact angle $\theta$ as~\cite{briant:04}
\begin{equation}
\phi_1=2 \beta \tau_w \sqrt{2p_c \kappa} \;
\mathrm{sign} \left( \frac{\pi}{2} - 
\theta \right) \sqrt{\cos\frac{\alpha}{3}\left(1-\cos\frac{\alpha}{3}\right)}
\label{eq:phi1}
\end{equation}
where $\alpha=\mathrm{cos}^{-1}(\sin^2\theta)$ and the function
$\mathrm{sign}$ returns the sign of its argument.

\Eq{eq:cahn1} is used to constrain the density derivative for sites on
a flat part of the substrate. However, no such exact results are
available for sites at edges or corners. We work on the principle that
the wetting angle at such sites should be constrained as little as
possible so that, in the limit of an increasingly fine mesh, it is
determined by the contact angle of the neighbouring flat surfaces.

\begin{figure}
\begin{center}
\epsfig{file=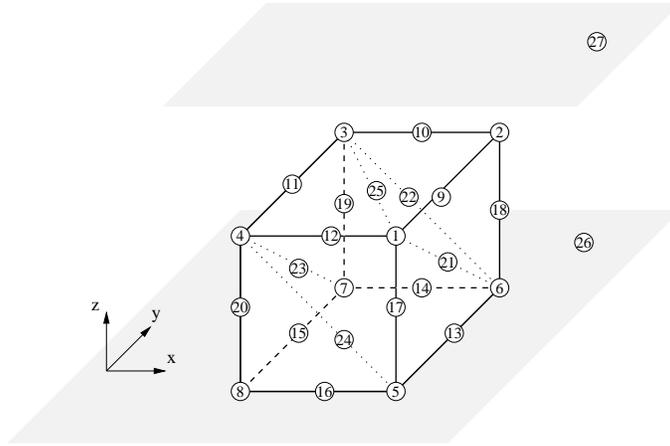,width=9cm}
\end{center}
\caption{Sketch of a post on a substrate. Encircled numbers label
  sites in different topological positions. Labels $26$ and $27$
  denote sites on the bottom ($z=z_{min}$) and the top ($z=z_{max}$)
  of the domain respectively.}
\label{fig:postmask}
\end{figure}

For edges (labels $9-12$ in \fig{fig:postmask}) and corners (labels
$1-4$) at the top of the post each site has $6$ neighbours on the
computational mesh. Therefore these sites can be treated as bulk
sites.

At bottom edges where the post abuts the surface (labels $13-16$ in
\fig{fig:postmask}) density derivatives in the two directions normal
to the surface (e.g. $x$ and $z$ for sites labeled $13$) are
calculated using
\begin{equation}
\partial_z n = \partial_{x/y} n 
             = - \frac{1}{\sqrt{2}} \frac{\phi_1}{\kappa}
\label{eq:cahn2}
\end{equation}
where the middle term constrains the density derivative in the
appropriate direction $x$ or $y$.

At bottom corners where the post joins the surface (labels $5-8$ in
\fig{fig:postmask}) density derivatives in both the $x$ and $y$ directions
are known. Therefore these sites are treated as planar sites.

\subsection{Velocity boundary conditions}

We impose a no-slip boundary condition on the velocity by determining
the missing fields which fulfill the no-slip condition given by
\eq{lb:velocity} with $\u=0$. This does not uniquely determine the
$f_i$'s. For most of the cases (i.e.  $1-20$) arbitrary choices guided
by symmetry are used to close the system. This is no longer possible
for sites $21-27$ where four asymmetrical choices are
available. Selecting one of those solutions or using a simple
algorithm which chooses one of them at random each time step leads to
very comparable and symmetrical results. Hence we argue that an
asymmetrical choice can be used. Possible conditions, which are used
in the results reported here, are listed in appendix A.

The conservation of mass is ensured by setting a suitable rest field,
$f_0$, equal to the difference between the density of the missing
fields and the one of the fields entering the solid after collision.

In a hydrodynamic description of wetting contact line slip must be
introduced in some way. As with other phase field models slip appears
naturally within the lattice Boltzmann framework. The mechanism
responsible for the slip is evaporation and condensation of the fluid
because of a non-equilibrium curvature of the contact
line~\cite{briant:04,jacqmin:00}.

% ============================================================
\section{Results}
\label{sec:results}

\begin{figure}
\begin{center}
\epsfig{file=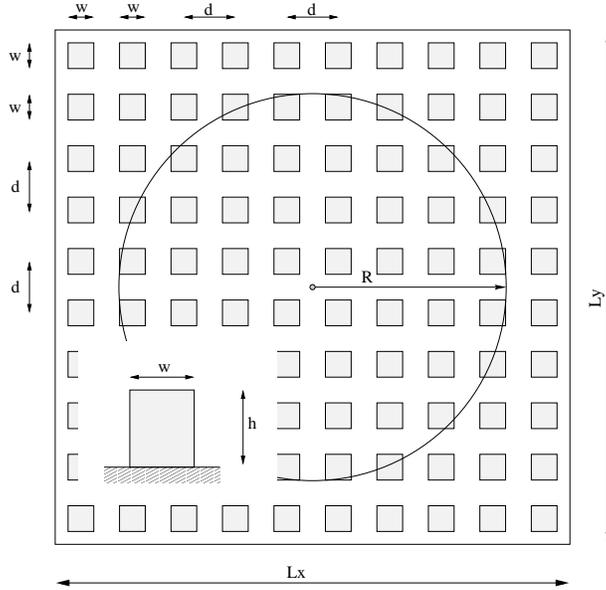,width=8cm}
\end{center}
\caption{Plan view of the substrate. Shaded areas are posts.}
\label{fig:substrate}
\end{figure}

We consider the superhydrophobic behaviour of droplet spreading on a
substrate patterned by square posts arranged as in
\fig{fig:substrate}. The size of the domain is $L_x \times L_y 
\times L_z = 80 \times 80 \times 80$ and the height, spacing
and width of posts are $h=5$, $d=8$ and $w=4$ respectively. A
spherical droplet of radius $R=30$ is initially centered within the
domain and just touches the post tops. The contact angle
$\theta^{eq}=110^o$ is set on every substrate site. The surface
tension and the viscosity are tuned by choosing parameters
$\kappa=0.002$ and $\tau=0.8$ respectively. The liquid density $n_l$
and gas density $n_g$ are set to $n_l=4.128$ and $n_g=2.913$ and the
temperature $T=0.4$.

%In what follows we will express density, velocity, time, force and
%energy as dimensionless numbers where respectecively $n^*=n \cdot
%n_l^{-1}$, $u^*=u \cdot \nu n_l \gamma^{-1}$, $t^*=t \cdot \gamma
%\nu^{-1} n_l^{-1} \Phi^{-1}$, $F^* = F \cdot \gamma^{-2} \nu^{2} 
%n_l^{2} \Phi$ and $E^*=E \cdot \nu^{-2} n_l^{-1} \Phi^{-1}$.

%We choose $T=0.4$ which corresponds to a liquid density $n_l=1$ and
%gas density $n_g=0.7$. They are unphysically small. This is necessary
%to achieve a stable simulation and must be taken into account by
%renormalising the time scale. {\em Is that a good way of saying it?
%Because later, the time scale is not renormalise.}.

Simulation and physical parameters can be related by choosing a length
scale $\Delta r$, a time scale $\Delta t$ and a mass scale $\Delta
m$. A simulation parameter with dimensions
$[L]^{n_1}[T]^{n_2}[M]^{n_3}$ is multiplied by $\Delta r^{n_1}\Delta
t^{n_2}\Delta m^{n_3}$ to give the physical value. For example,
considering a $1$ mm droplet of a fluid of kinematic viscosity
$\nu=3\cdot10^{-5}$ m$^2$s$^{-1}$ and surface tension $1 \cdot
10^{-3}$ Nm$^{-1}$ leads to $\Delta r=1.7\cdot10^{-5}$ m, $\Delta
t=9.3\cdot10^{-7}$ s, $\Delta m=1.6\cdot10^{-12}$ kg. That implies
$n_l=1.4\cdot 10^{3}$ kgm$^{-3}$. However, as with all diffuse
interface models the liquid-gas density difference is unphysically
small and the width of the interface is too large. This must be taken
into account by renormalising the time scale.

\subsection{Equilibrium states}

\begin{figure}
\begin{center}
\begin{tabular}{m{1cm}m{5cm}m{0.5cm}m{5cm}}
(a) & \epsfig{file=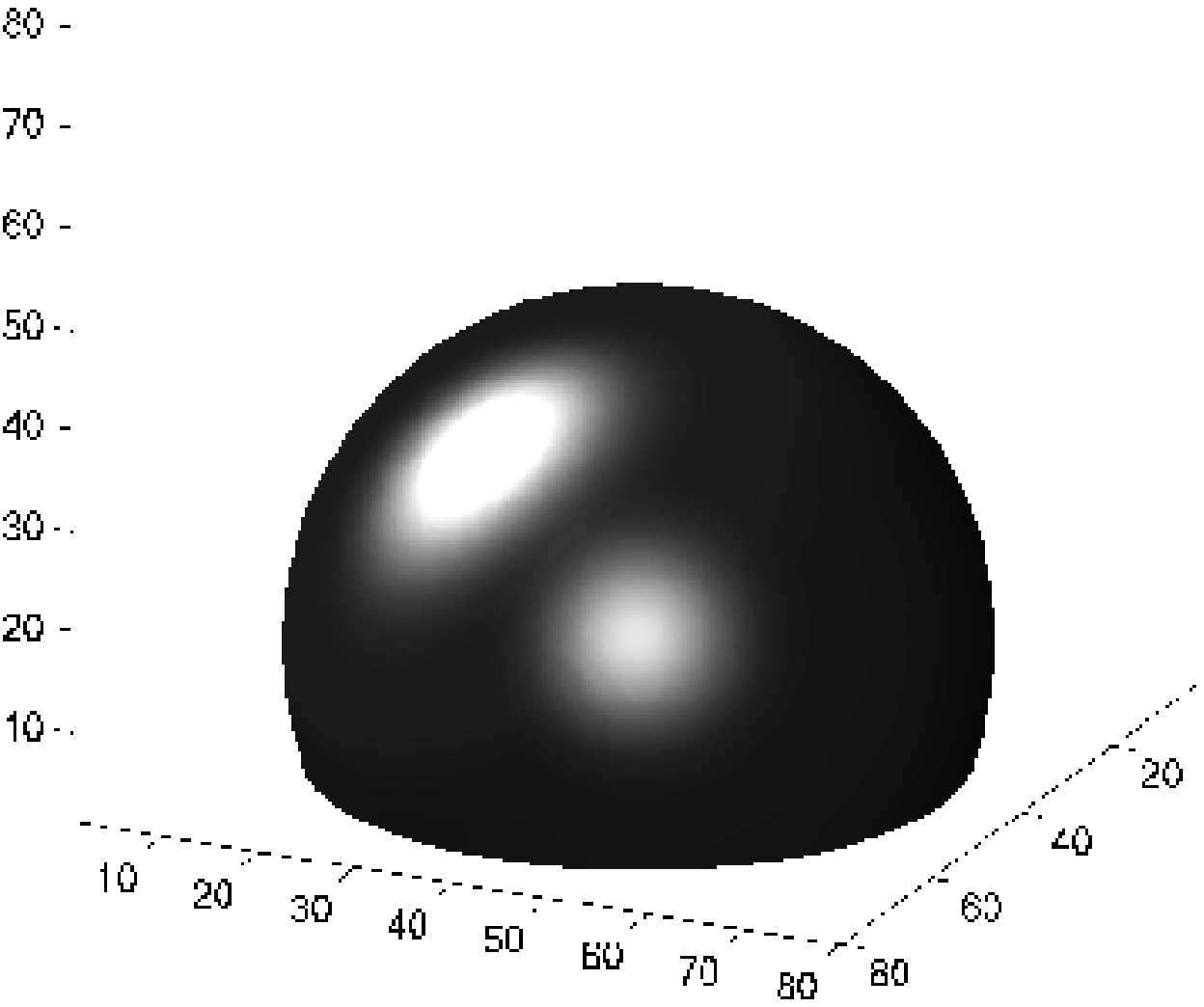,width=5cm} &&
  \epsfig{file=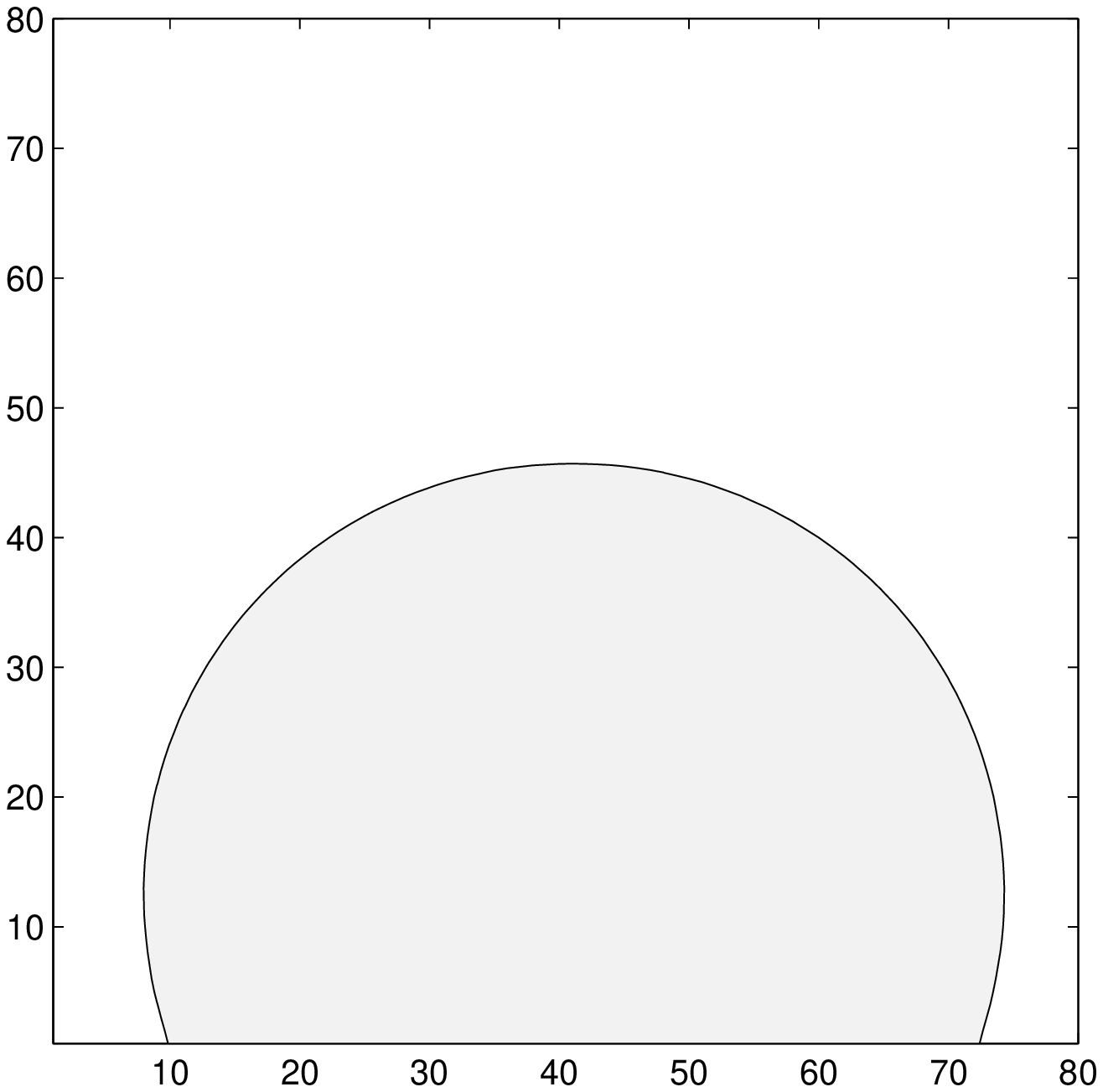,width=5cm} \\
(b) & \epsfig{file=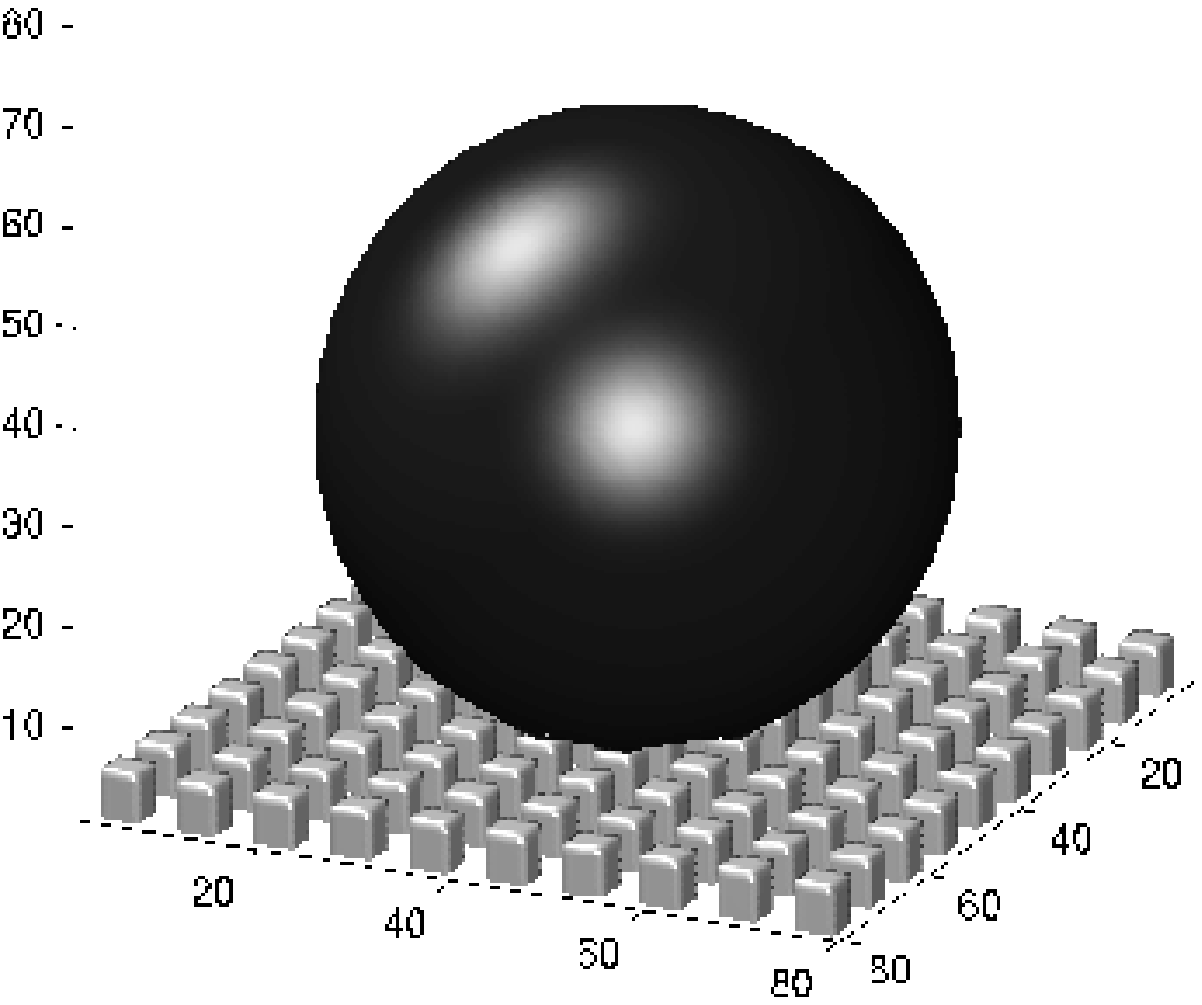,width=5cm} &&
  \epsfig{file=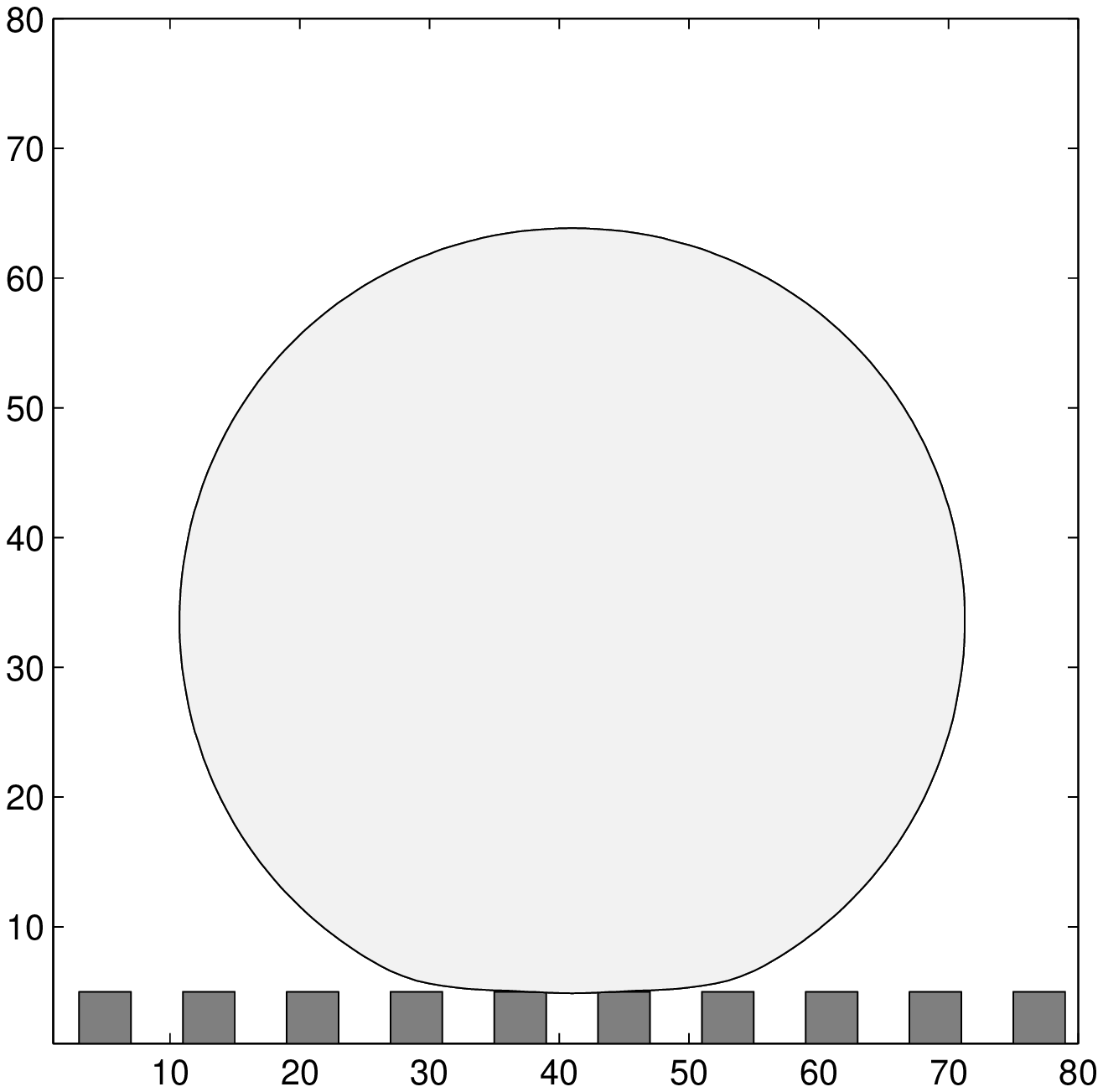,width=5cm} \\
(c) & \epsfig{file=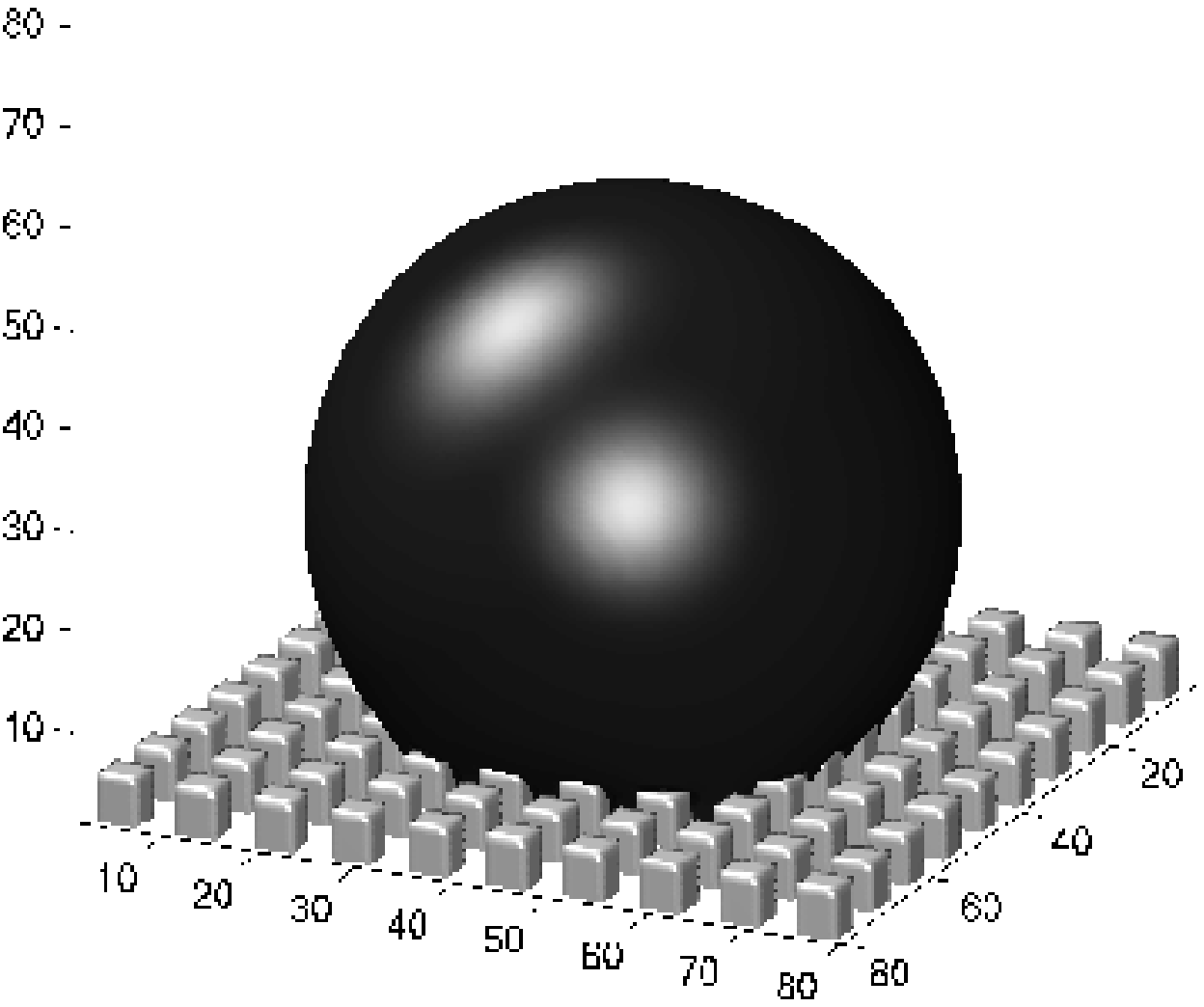,width=5cm} &&
  \epsfig{file=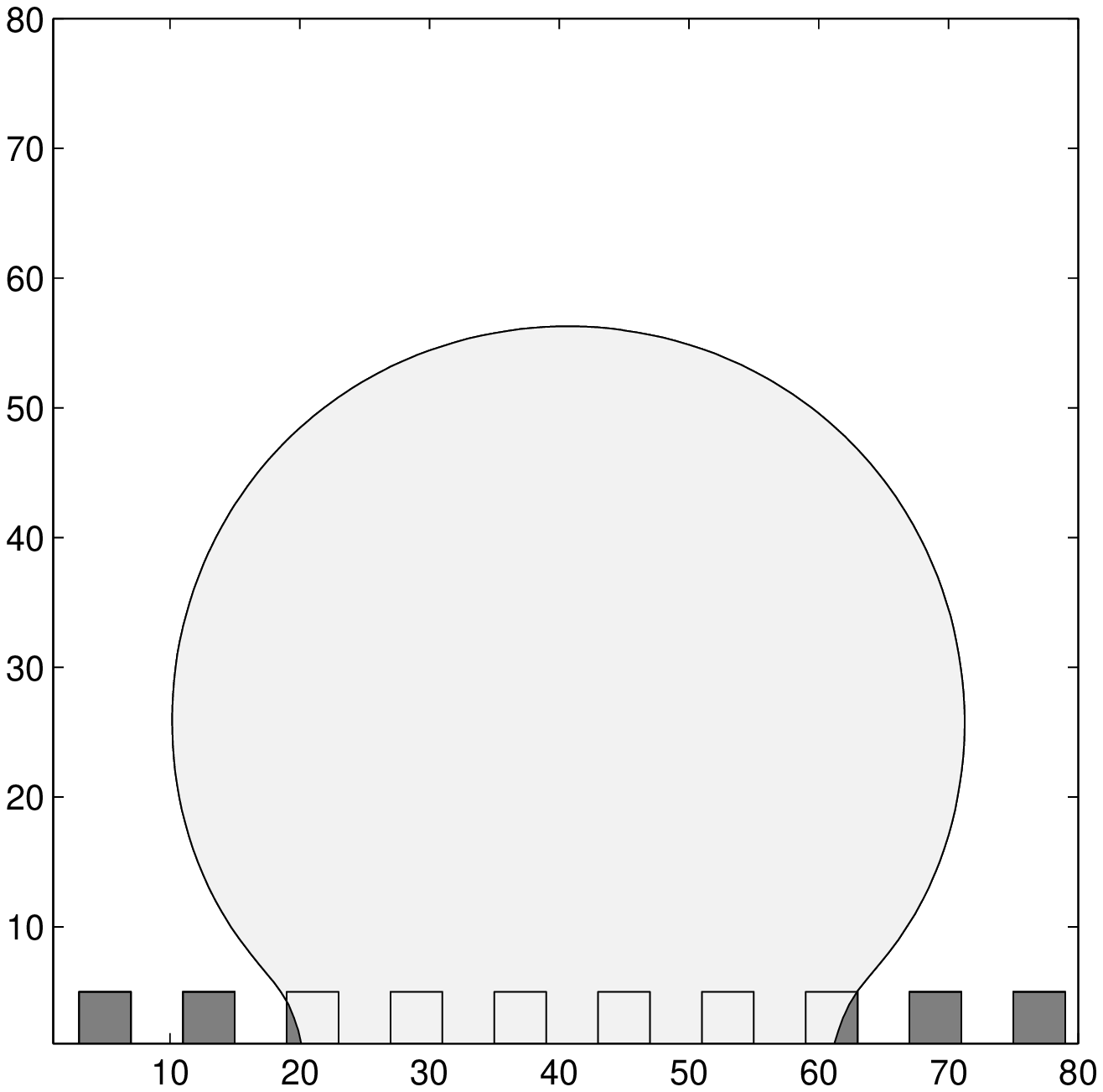,width=5cm} \\
\end{tabular}
\end{center}
\caption{Final states of a spreading droplet. The right column reports 
vertical cuts across the centre of the drop. (a) The substrate is flat
and homogeneous. (b) The substrate is decorated with posts and the
initial velocity of the droplet is zero. (c) Same geometry as (b) but
the droplet reaches the substrate with a velocity $u_z=-0.01$.}
\label{fig:results1}
\end{figure}

\Fig{fig:results1} shows the final state attained by the droplet for 
different substrates or initial conditions. For comparison
\fig{fig:results1}(a) shows a planar substrate. The equilibrium
contact angle is $\theta^{flat} = 110^o = \theta^{eq}$ as
expected~\cite{dupuis:03c}. In
\fig{fig:results1}(b) the substrate is patterned and the initial
velocity of the drop is zero. Now the contact angle is $\theta^s =
156^o$, a demonstration of superhydrophobic
behaviour. \Fig{fig:results1}(c) reports an identical geometry but a
drop with an initial impact velocity. Now the drop is able to collapse
onto the substrate and the final angle is $\theta^c = 130^o$. These
angles are compatible with the ones reported in~\cite{oner:00}. 

Superhydrophobicity occurs over a wide range of $d$, the distance
between the posts. For suspended drops of $R=30$ and $d \gtrsim 12$
the drop resides on a single post and the contact angle is
$170^o$. For $d \lesssim 12$ the contact angle lies between $148^o$
and $156^o$ with the range primarily due to the commensurability
between drop radius and post spacing.

\subsection{Kinetics of the suspended to collapsed transition}

We now investigate the kinetics of the transition between the
suspended and collapsed droplet states. For the parameter values used
in the simulations presented in fig.~\ref{fig:results1} the state with
the drop suspended on the posts has a slightly higher free energy than
the collapsed state. However as the drop penetrates the grooves the
area of the contact between liquid and solid increases. Because the
substrate is hydrophobic this creates a free energy barrier hindering
the transition from the suspended to the collapsed states. Work must
be provided, by an impact velocity or gravity say, to allow the
transition to proceed~\cite{lafuma:03,he:03}.

\begin{figure}
\begin{center}
\begin{tabular}{ccc}
$t_1=70000$ & $t_2=75000$ & $t_3=100000$ \\
\epsfig{file=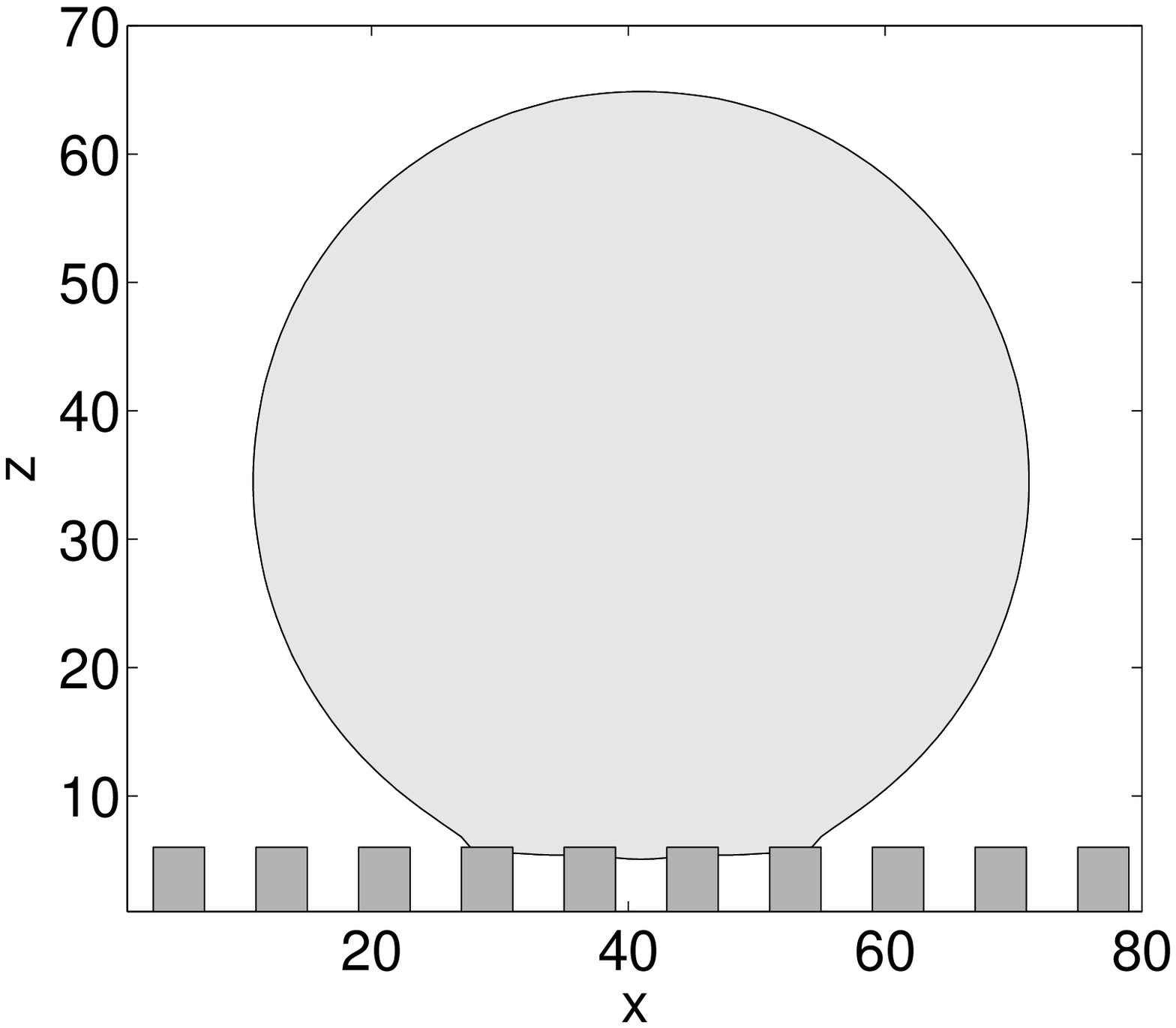,width=4cm} &
\epsfig{file=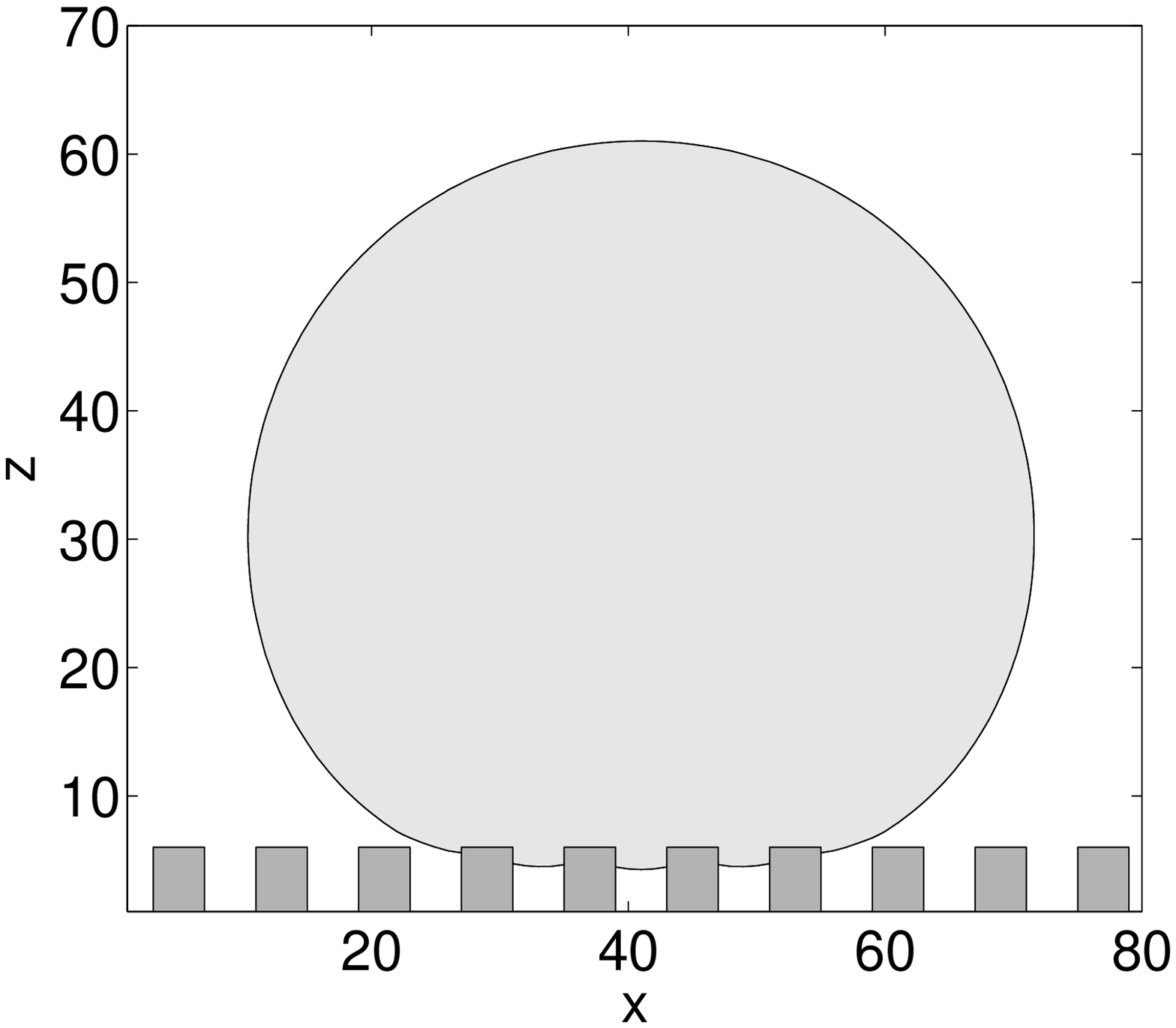,width=4cm} &
\epsfig{file=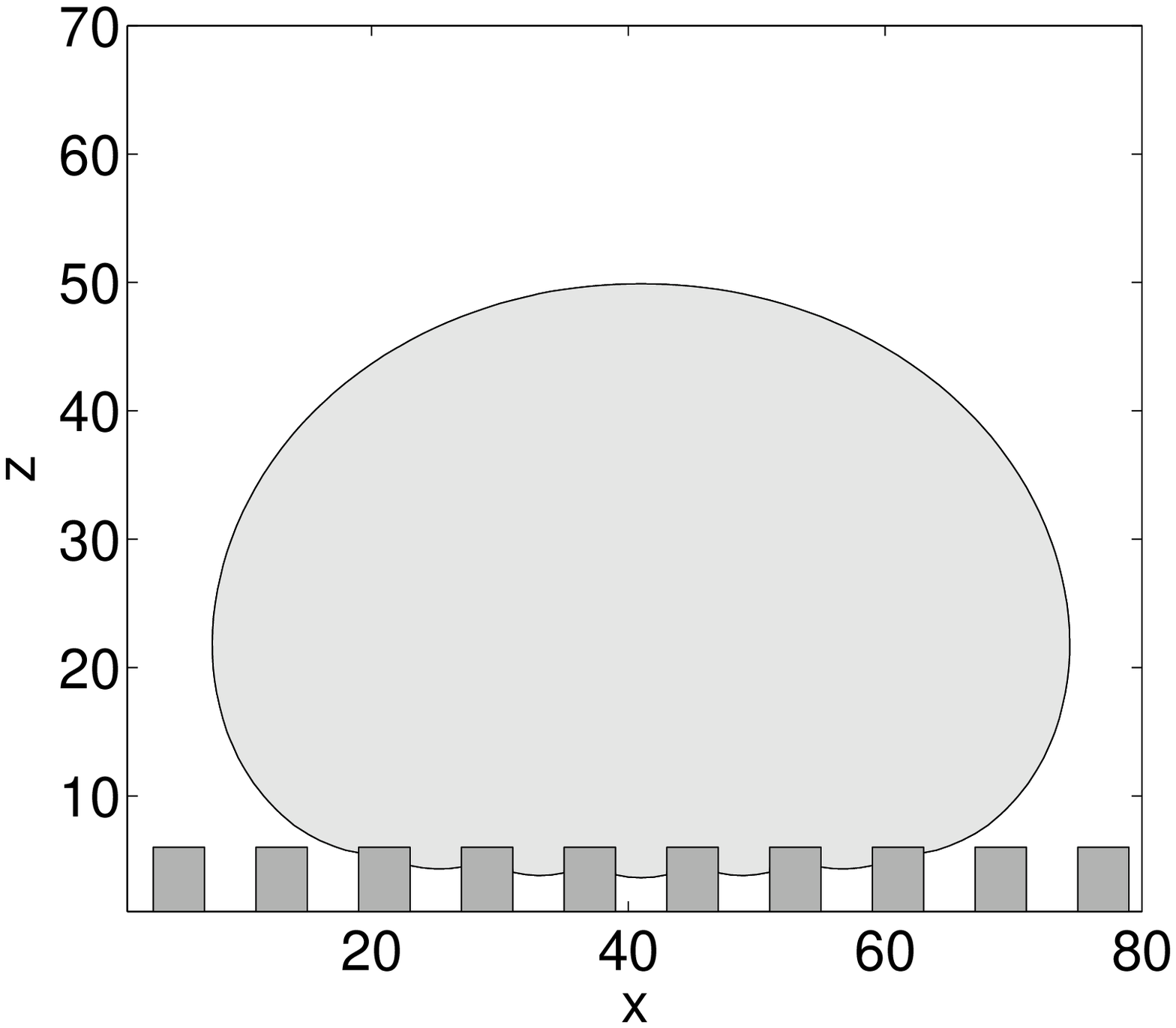,width=4cm} \\
$t_4=124000$ & $t_5=126000$ & $t_6=129500$ \\
\epsfig{file=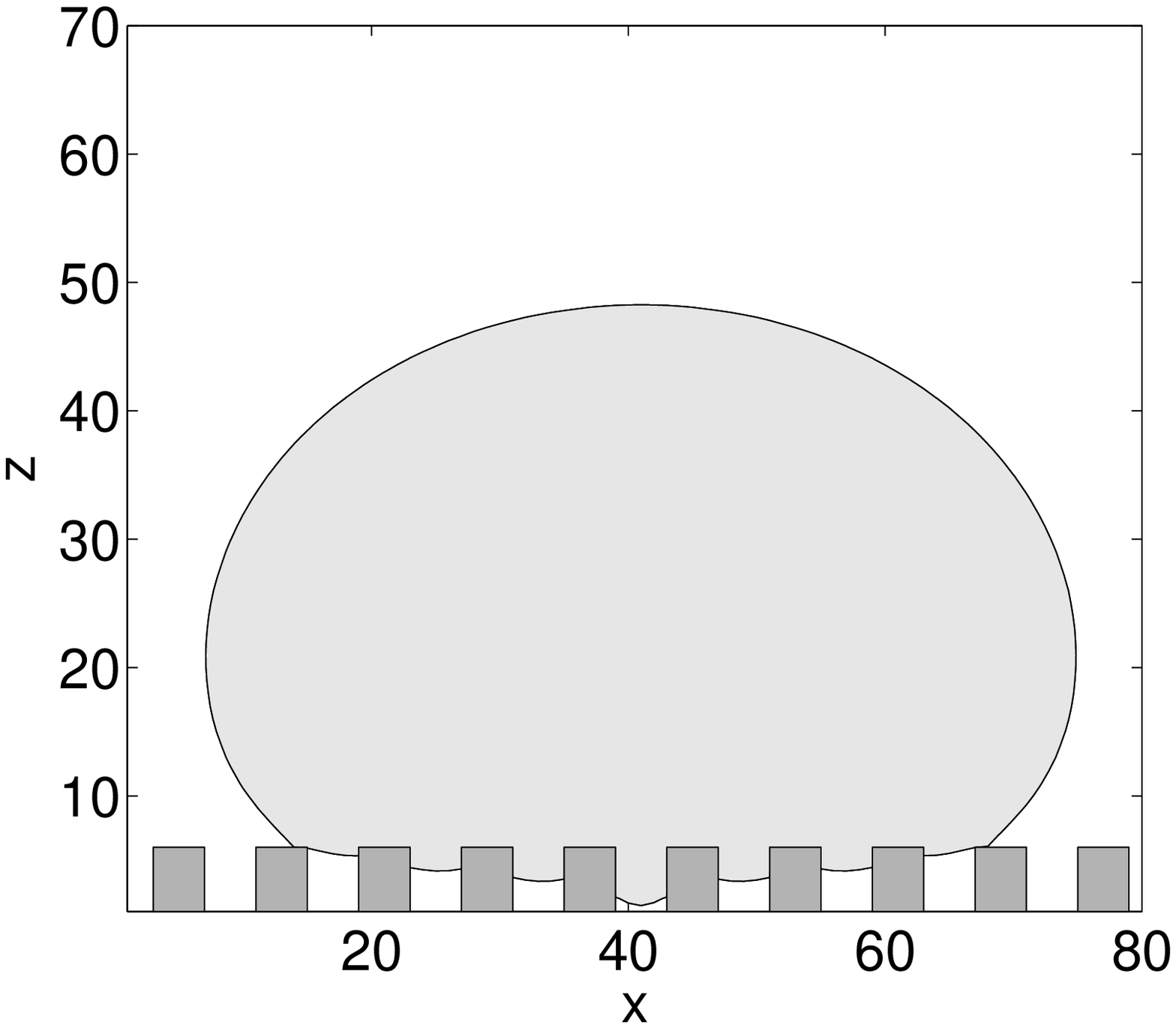,width=4cm} &
\epsfig{file=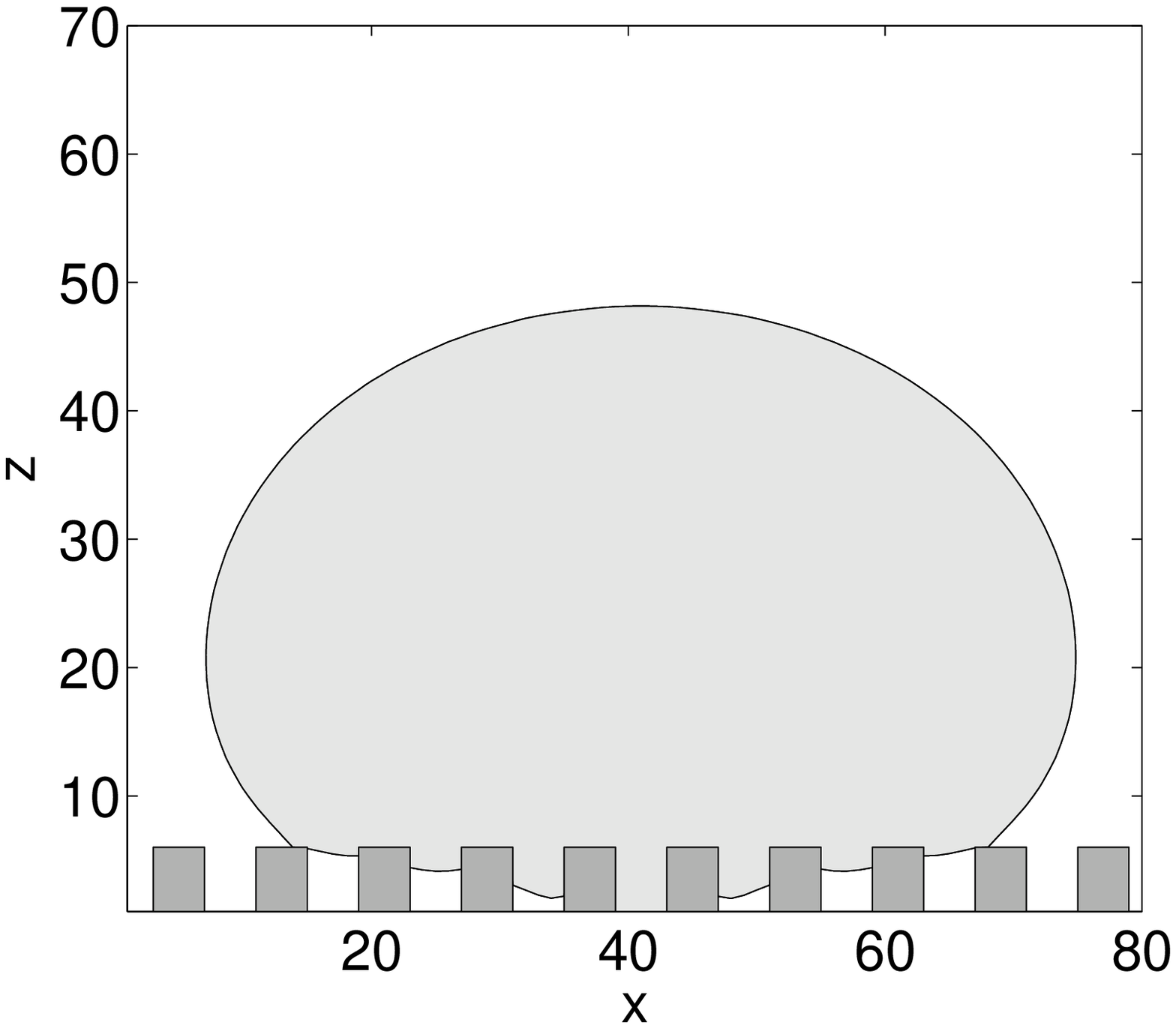,width=4cm} &
\epsfig{file=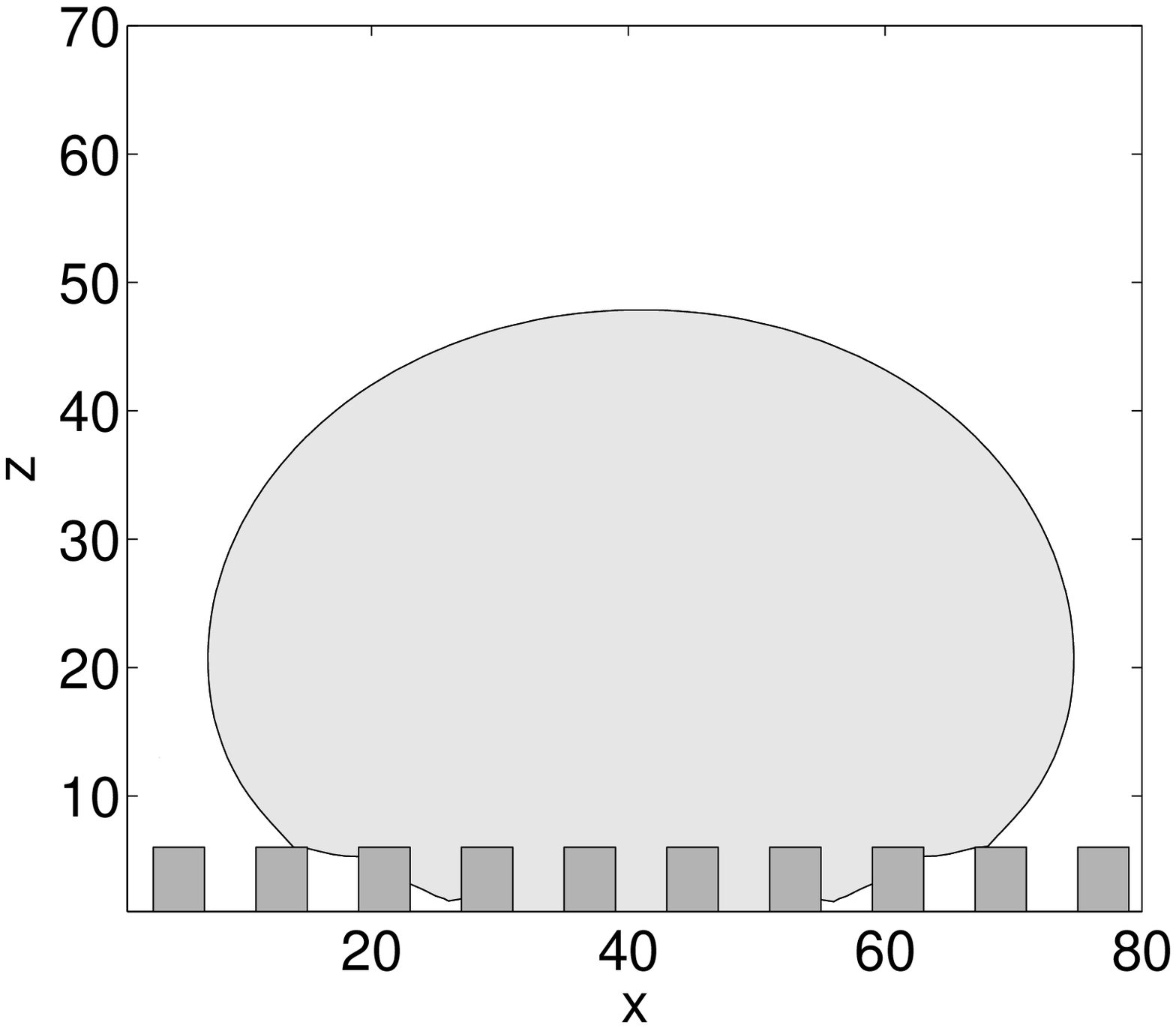,width=4cm} \\
$t_7=138000$ & $t_8=140000$ & $t_9=300000$ \\
\epsfig{file=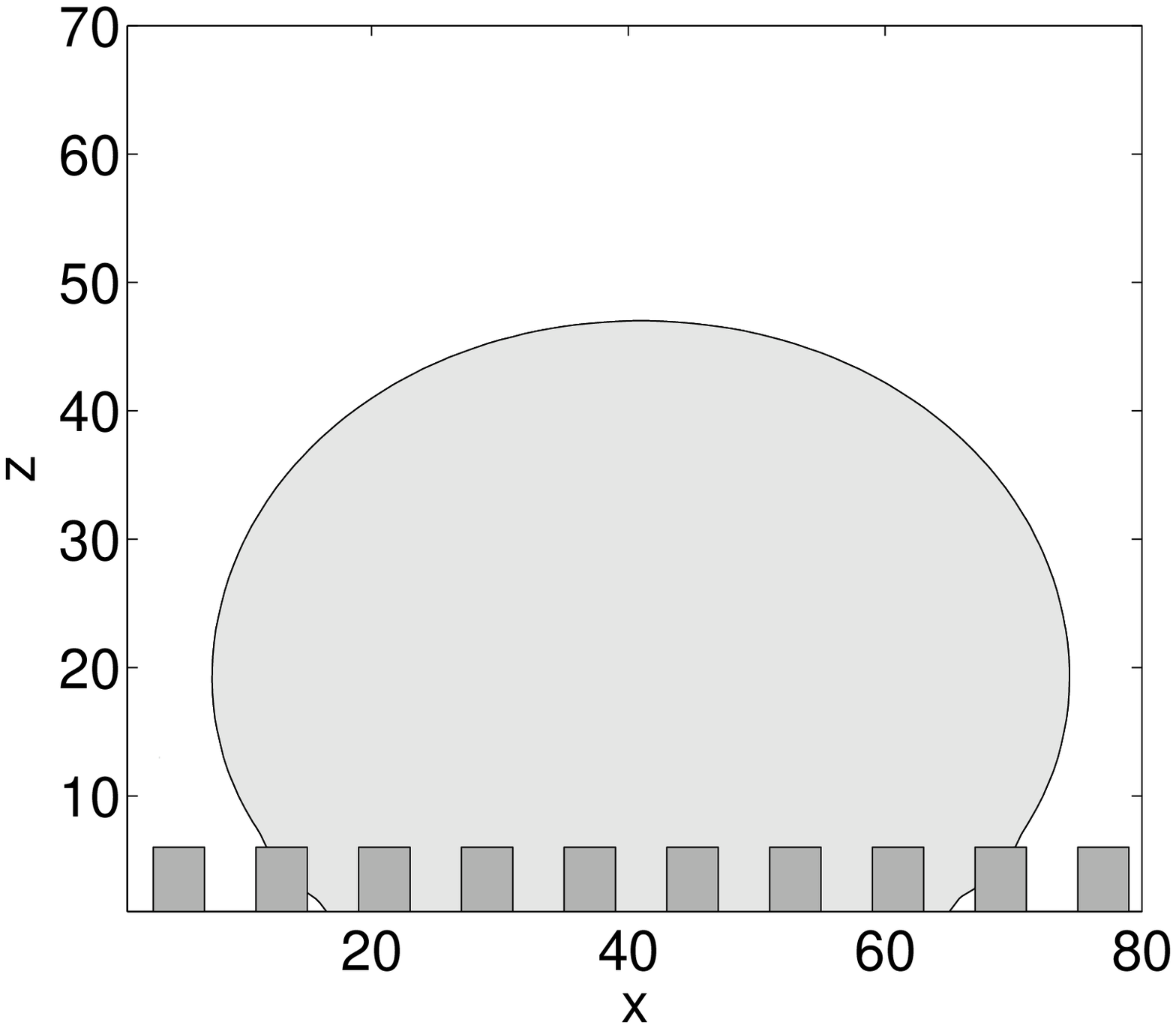,width=4cm} &
\epsfig{file=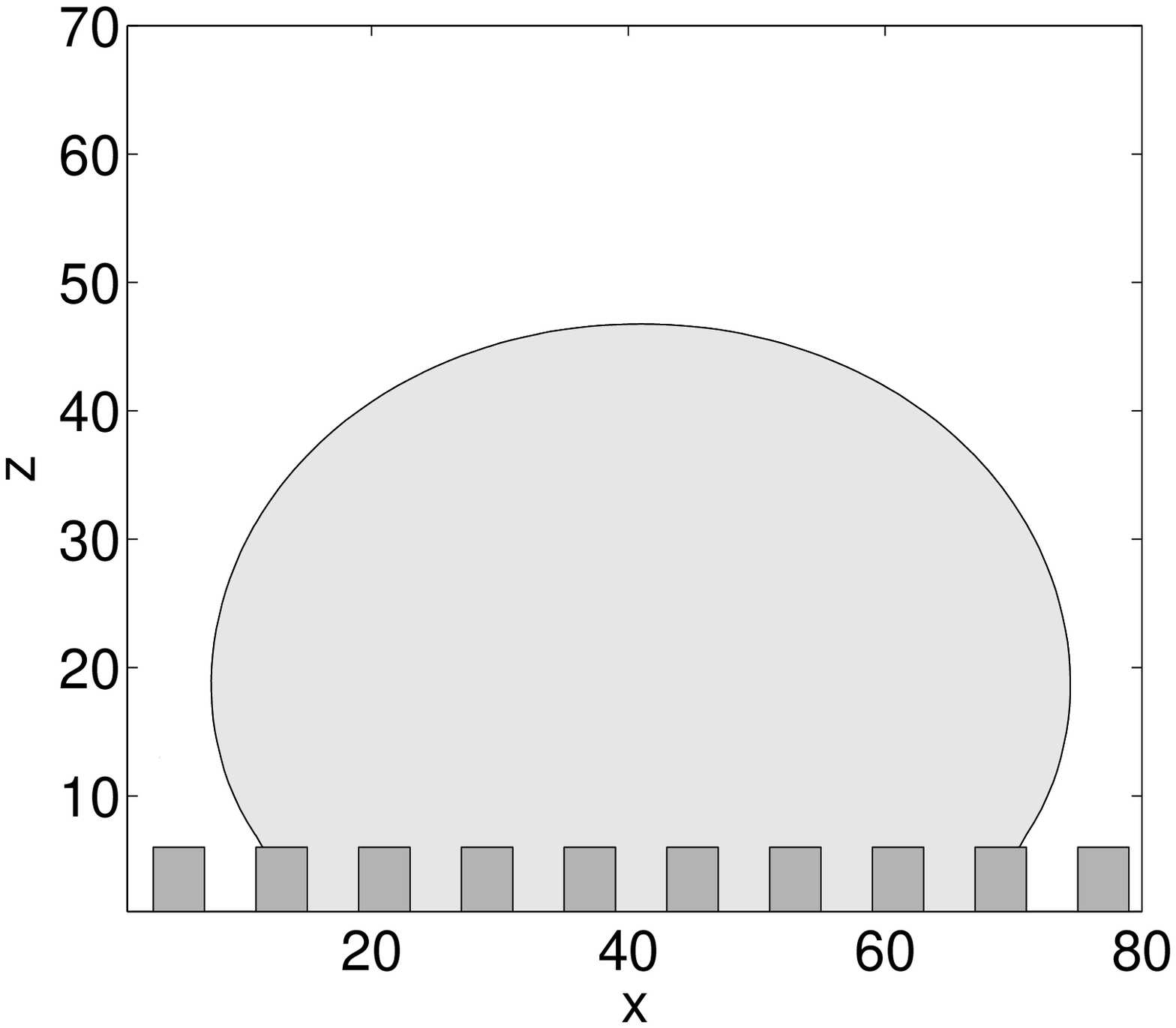,width=4cm} &
\epsfig{file=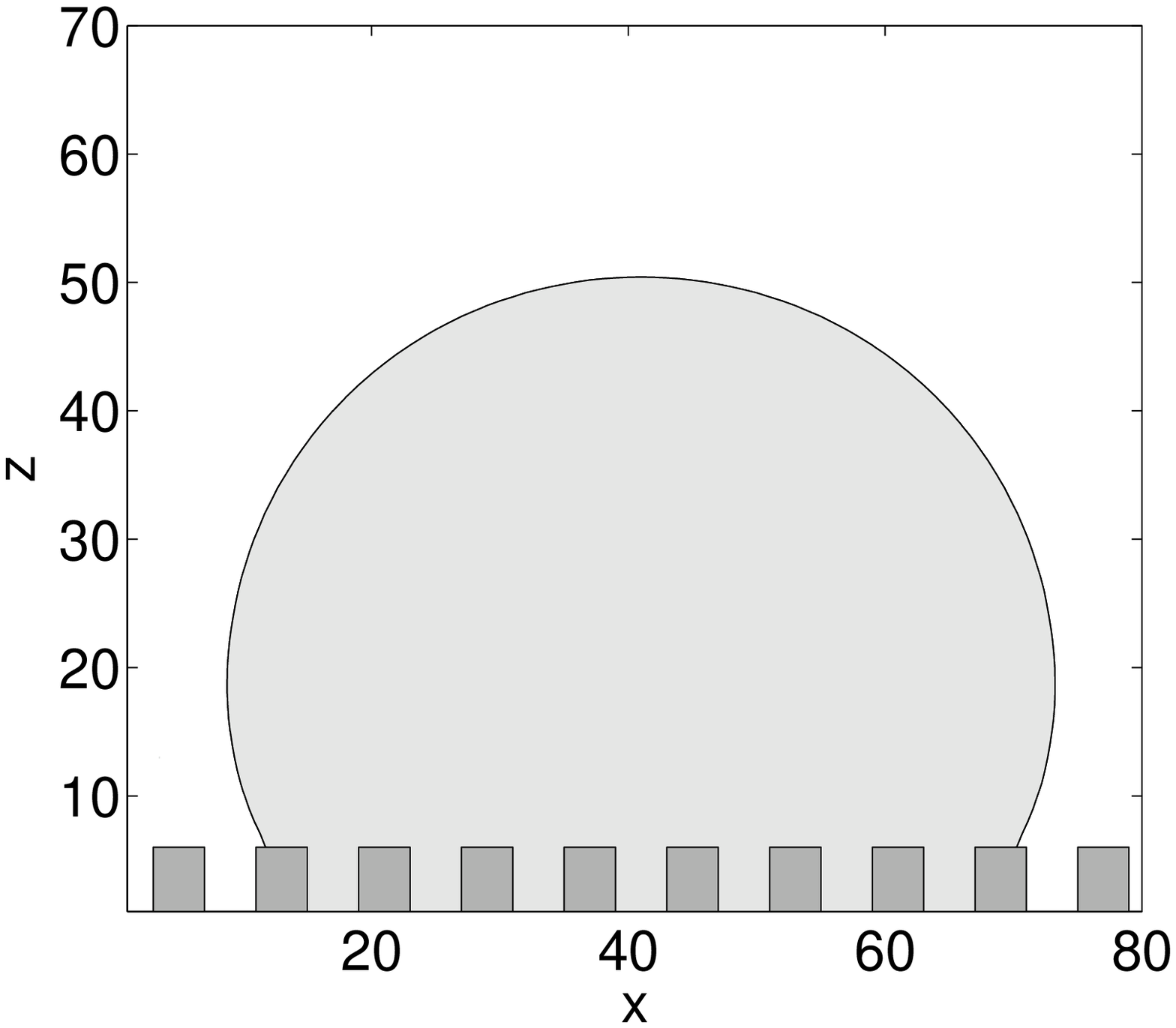,width=4cm} \\
\end{tabular}
\end{center}
\caption{Transition from a suspended to a collapsed state. The 
gravity field $F_z=5\cdot10^{-7}$. These cuts are vertical cross
sections across the centre of the domain where the dark gray areas are
the posts and the pale ones are liquid regions.}
\label{fig:snapsE}
\end{figure}

We follow the transition pathway by considering a spherical droplet of
radius $R=30$ initially centered within the domain and just touching
the top of the posts. A gravitational field $F_z$ is turned on at time
$t=70000$, and turned off at $t=200000$. Fig.~\ref{fig:snapsE} shows
cross sections of the drop as it undergoes the transition from the
suspended to the collapsed state for $F_z=-5\cdot10^{-7}$. Note how the
substrate interstices are filled successively from the drop centre to
its edges.

\begin{figure}
\begin{center}
\begin{tabular}{cc}
\epsfig{file=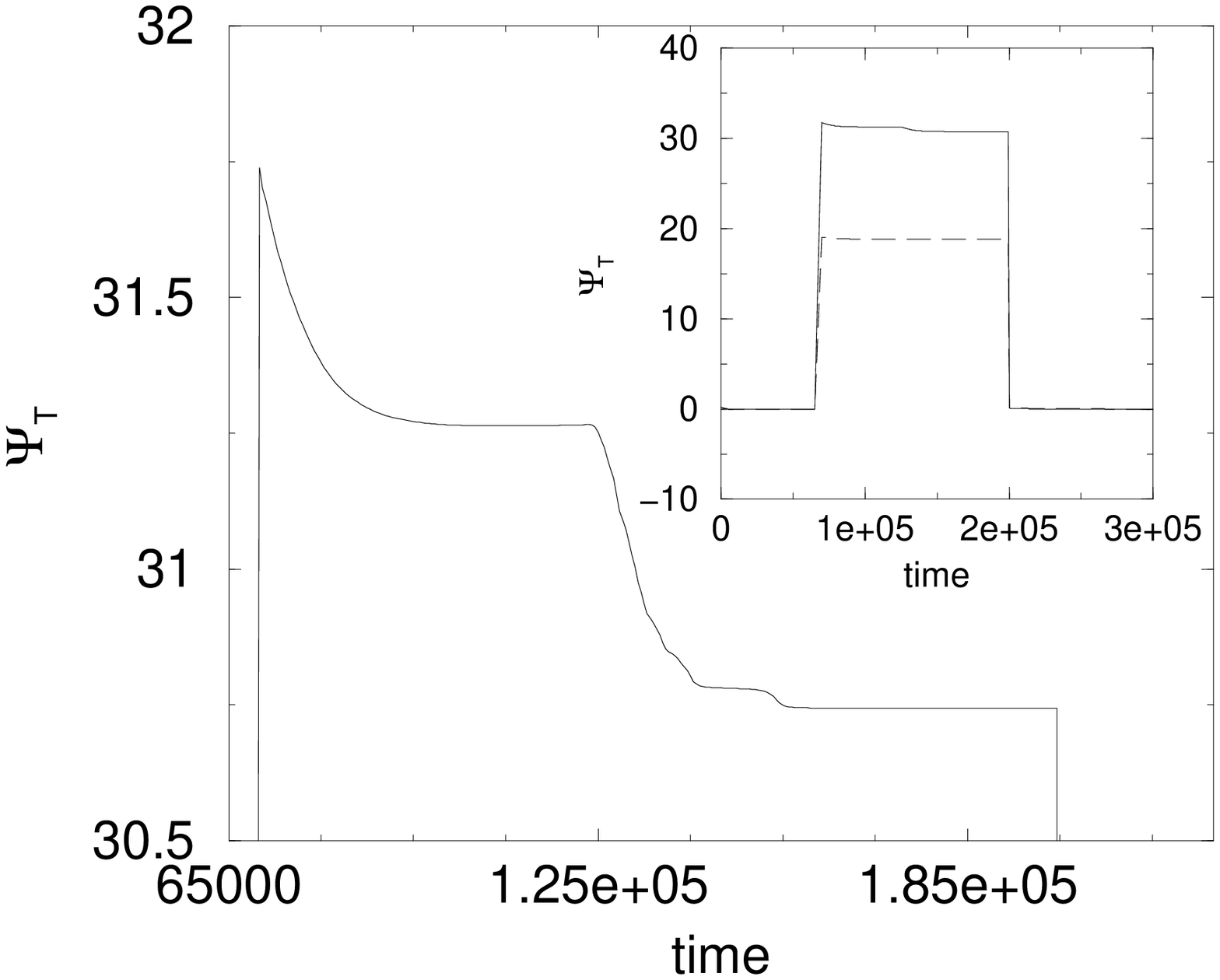,height=5cm} & \epsfig{file=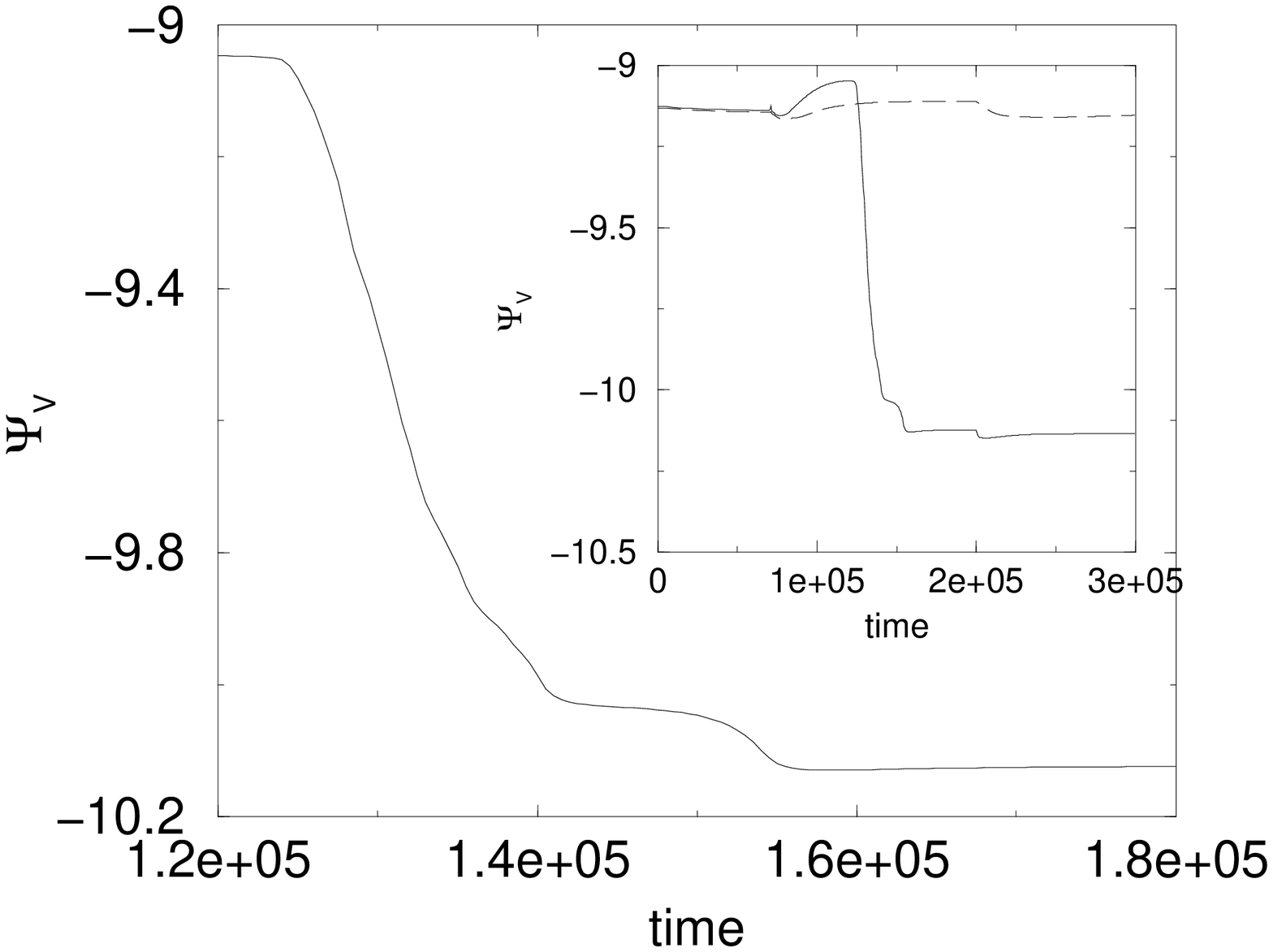,height=5cm} \\
\epsfig{file=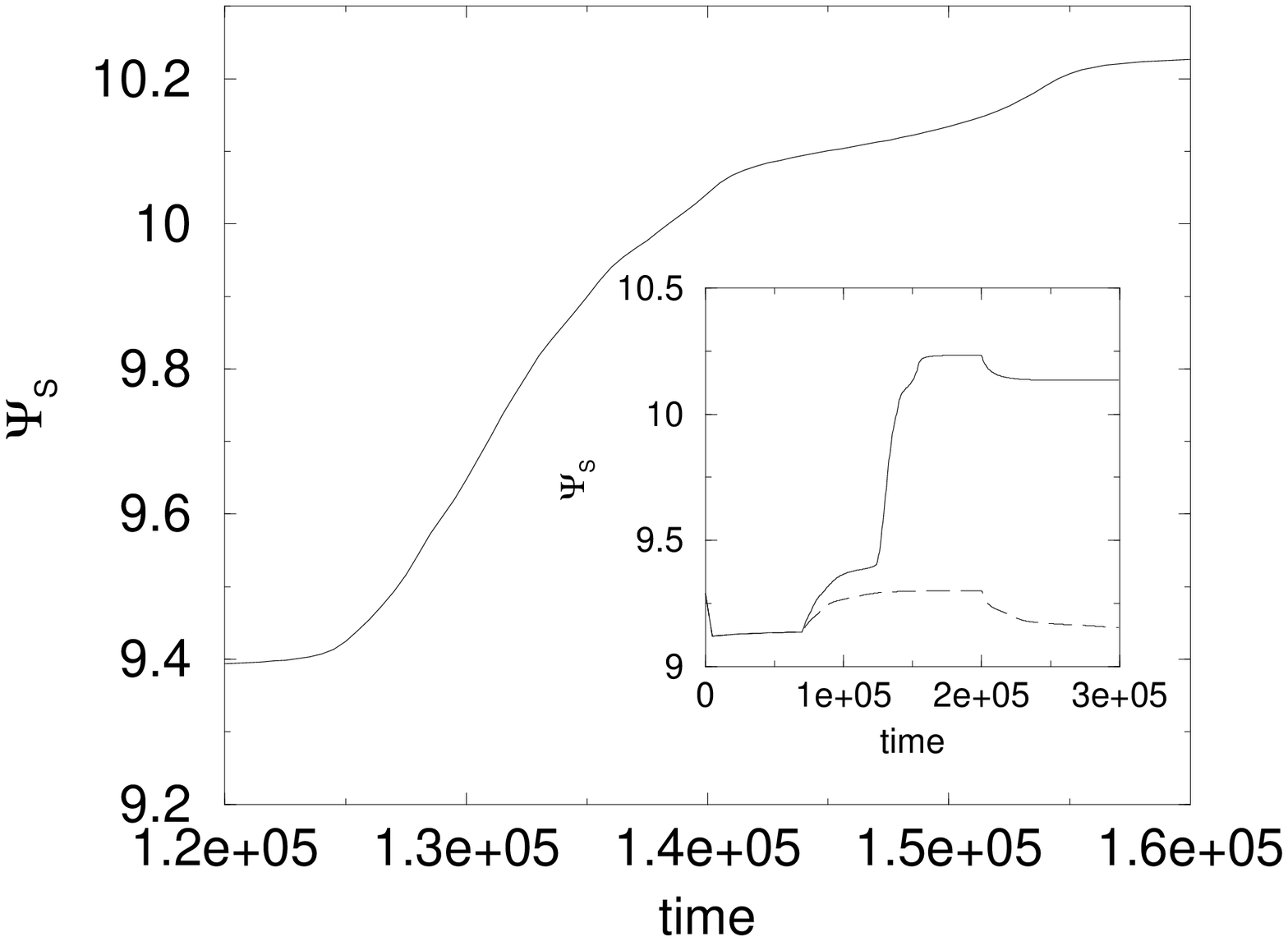,height=5cm} & \epsfig{file=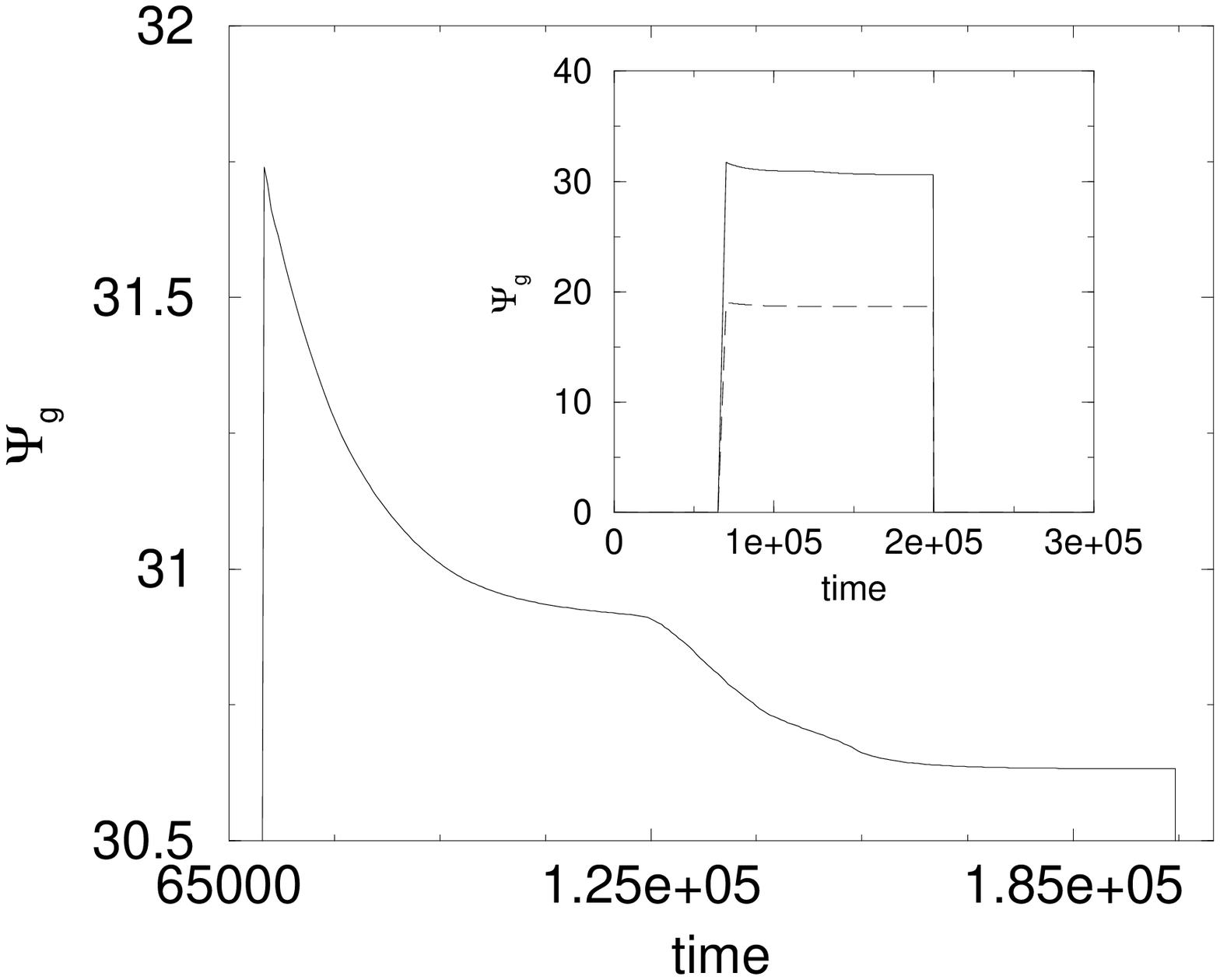,height=5cm} \\
\end{tabular}
\end{center}
\caption{Evolution of the free energies of superhydrophobic drops. A 
body force $F_z$ is turned on at $t=70000$ and turned off at
$t=200000$. Solid and dashed lines corresponds to free energies when
$F_z=-5\cdot10^{-7}$ and $F_z=-3\cdot10^{-7}$ respectively. The axes
are in simulation units with free energy chosen to be zero at
equilibrium. In each case the larger diagram highlights the free
energy and time regimes where the most interesting behaviour
occurs. The insets show the behaviour throughout the simulation.}
\label{fig:Energy}
\end{figure}

The time evolution of the total free energy of the system is presented
in fig.~\ref{fig:Energy} for two different values of $F_z$. We plot
$\Psi_{T}$ the total free energy; the volume $\Psi_{v}$ and surface
$\Psi_{s}$ contributions to the free energy which arise from the first
and second integral in (1) respectively and the contribution from the
gravitational force
\begin{equation}
\Psi_{g} = \int_V nzF_z \; dV.
\end{equation}

The solid lines in fig.~\ref{fig:Energy} correspond to the snapshots
in fig.~\ref{fig:snapsE}. After gravity is switched on the drop is
pushed down and hence $\Psi_{g}$ decreases. On the other hand
$\Psi_{s}$ increases as the surface is covered by liquid rather than
by gas corresponding to a free energy gain for a hydrophobic
surface. $\Psi_{v}$ also initially increases as the interface is
deformed. Once the drop touches the bottom of the interstices
$\Psi_{v}$ drop sharply because parts of the interface have just
vanished. At the same time $\Psi_{s}$ grows significantly because a
part of the surface is now in contact with the liquid rather than with
air. At $t=200000$, the gravitational field is turned off. The
squeezed droplet dewets to recover a spherical shape and $\Psi_{s}$
decreases. The total free energy is slightly smaller in the collapsed
state.

The figures show that both the surface and the volume energies
increase during the transition. Hence, as the total free energy must
decrease, the transition could not occur without the addition of a
gravitational force. The decrease in the gravitational free energy
offsets the increase in the surface and volume terms allowing the
transition to proceed. Essentially the addition of $F_z$ removes the
barrier which the total free energy must surmount.

A lower gravitational field, $F_z=3\cdot10^{-7}$, does not allow the
drop to overcome the free energy barrier and it remains in the
suspended state as shown by the dashed lines in fig.~\ref{fig:Energy}.

For example, considering again a $1$ mm droplet of a fluid of
kinematic viscosity $\nu=3\cdot10^{-5}$ m$^2$s$^{-1}$ and surface
tension $1 \cdot 10^{-3}$ Nm$^{-1}$, $F_z=5\cdot10^{-7}$ would
correspond to 9.7 ms$^{-2}$ in physical units.

% ============================================================
\section{Conclusion}
\label{sec:conclusion}

We have used a lattice Boltzmann approach to solve the equations of
motion describing the dynamics of a drop on topologically patterned
substrate. The approach allows us to simulate the dynamic and
equilibrium properties of a drop with a given size, surface tension,
contact angle and viscosity. Because interfaces appear naturally
within the model it is particularly well suited to looking at the
behaviour of evolving drops.
 
In particular we have considered a droplet positioned on an array of
posts and shown that it is possible to reproduce the superhydrophobic
behaviour seen in experiments. The ability of the algorithm to follow
the drop kinetics has enabled us to investigate the transitions
between the suspended state where the drop lies on top of the posts
and the collapsed state where it fills the spaces between them. We
find that the substrate interstices are filled successively starting
from the drop centre.

There are many avenues open for further investigation. For example we
are currently investigating how drops move across hydrophobic surfaces
to compare to recent experiments. It would also be interesting to
follow the spreading of drops on surfaces with topological
imperfections, a problem of concern in the quality of images formed in
ink-jet printing.

% ============================================================
\section*{Acknowledgment}

We thank the Oxford Supercomputing Centre for providing supercomputing
resources. AD acknowledges the support of the EC IMAGE-IN project
GR1D-CT-2002-00663.

\begin{appendix}

\section{Possible boundary conditions}

We list here the boundary conditions used to define missing
distribution functions, i.e. those that stream from positions outside
the simulation box.

\begin{center}

\tablehead{\hline Label (see fig.2) & \multicolumn{2}{c|}{Conditions} \\ \hline\hline}
\tabletail{\hline}
           
\begin{supertabular}{|c|ll|}
$1$ & $f_{13}$ & $=f_{14}$ \\ \hline
$2$ & $f_{7}$  & $=f_{8}$ \\ \hline
$3$ & $f_{9}$  & $=f_{10}$ \\ \hline
$4$ & $f_{11}$ & $=f_{12}$ \\ \hline

$5$ & $f_{5}$  & $=f_{6}$ \\
    & $f_{13}$ & $=(f_3-f_4-f_1+f_2)/2+f_9+f_{14}-f_{10}$ \\
    & $f_{11}$ & $=(f_1-f_2)/2-f_{9}+f_{10}+f_{12}$ \\
    & $f_7$    & $=(-f_3+f_4)/2+f_8-f_9+f_{10}$ \\ \hline   

$6$ & $f_5$    & $= f_6$ \\
    & $f_{13}$ & $= (f_3-f_4)/2-f_{11}+f_{12}+f_{14}$ \\
    & $f_9$    & $= (f_1-f_2)/2-f_{11}+f_{10}+f_{12}$ \\
    & $f_7$    & $= (-f_1+f_2-f_3+f_4)/2+f_8+f_{11}-f_{12}$ \\ \hline

$7$ & $f_5$     & $= f_6$ \\
    & $f_{11}$  & $= (f_3-f_4)/2-f_{13}+f_{12}+f_{14}$ \\
    & $f_9$     & $= (f_1-f_2-f_3+f_4)/2+f_{13}-f_{14}+f_{10}$ \\
    & $f_7$     & $= (-f_1+f_2)/2+f_8-f_{13}+f_{14}$ \\ \hline

$8$ & $f_5$     & $= f_6$ \\ 
    & $f_{11}$  & $= (f_3-f_4+f_1-f_2)/2+f_7-f_8+f_{12}$ \\
    & $f_9$     & $= (-f_3+f_4)/2-f_7+f_8+f_{10}$ \\
    & $f_{13}$  & $= (-f_1+f_2)/2-f_7+f_8+f_{14}$ \\ \hline

$9$  & $f_{13}$ & $= f_{14} \quad ;  \quad f_7 = f_8$ \\ \hline
$10$ & $f_9$    & $= f_{10} \quad ;  \quad f_7 = f_8$ \\ \hline
$11$ & $f_9$    & $= f_{10} \quad ;  \quad f_{11} = f_{12}$ \\ \hline
$12$ & $f_{13}$ & $= f_{14} \quad ;  \quad f_{11} = f_{12}$ \\ \hline

$13$ & $f_5$     & $= f_6$ \\
    & $f_1$     & $= 2(-f_{10}+f_9+f_{11}-f_{12})+f_2$ \\
    & $f_{13}$  & $= (f_3-f_4)/2 - f_{11} + f_{12} + f_{14}$ \\
    & $f_7$     & $= (-f_3+f_4)/2 + f_8 - f_9 + f_{10}$ \\ \hline

$14$ & $f_5$     & $= f_6$ \\
    & $f_9$     & $= (f_1-f_2)/2-f_{11}+f_{10}+f_{12}$ \\
    & $f_7$     & $= (-f_1+f_2)/2+f_8-f_{13}+f_{14}$ \\
    & $f_3$     & $= 2(-f_{12}+f_{11}+f_{13}-f_{14})+f_4$ \\ \hline

$15$ & $f_5$     & $= f_6$ \\
    & $f_2$     & $= 2(-f_{14}+f_7+f_{13}-f_8)+f_1$ \\
    & $f_{11}$  & $= (f_3-f_4)/2 - f_{13} + f_{12} + f_{14}$ \\
    & $f_9$     & $= (-f_3+f_4)/2 + f_8 - f_7 + f_{10}$ \\ \hline

$16$ & $f_5$     & $= f_6$ \\
    & $f_{11}$  & $ = (f_1-f_2)/2-f_9+f_{10}+f_{12}$ \\
    & $f_{13}$  & $= (-f_1+f_2)/2-f_7+f_8+f_{14}$ \\
    & $f_4$     & $= 2(-f_{10}+f_7+f_9-f_8)+f_3$ \\ \hline

$17$ & $f_{10}$   & $= f_9      \quad ;  \quad f_{13} = f_{14}$ \\ \hline
$18$ & $f_7$      & $= f_8    \quad ;  \quad f_{12} = f_{11}$ \\ \hline
$19$ & $f_9$      & $= f_{10} \;\; ;  \quad f_{14} = f_{13}$ \\ \hline
$20$ & $f_8$      & $= f_7    \quad ;  \quad f_{11} = f_{12}$ \\ \hline

$21$ & $f_1$    & $= f_2 \quad ;  \quad f_7 = f_8$ \\
    & $f_{12}$ & $= (-f_3+f_4)/2+f_{11}$ \\
    & $f_{13}$ & $= (-f_5+f_6)/2+f_{14}$ \\
    & $f_{10}$ & $= (f_3-f_4+f_5-f_6)/2+f_9$ \\ \hline

$22$ & $f_3$    & $= f_4 \quad ;  \quad f_7 = f_8$ \\
    & $f_9$    & $= (-f_5+f_6)/2+f_{10}$ \\
    & $f_{14}$ & $= (f_1-f_2+f_5-f_6)/2+f_{13}$ \\
    & $f_{12}$ & $= (-f_1+f_2)/2+f_{11}$ \\ \hline

$23$ & $f_2$    & $= f_1 \quad ;  \quad f_8 = f_7$ \\
    & $f_{11}$ & $= (f_3-f_4)/2+f_{12}$ \\
    & $f_{14}$ & $= (f_5-f_6)/2+f_{13}$ \\
    & $f_9$    & $= f_{10}+(-f_3+f_4-f_5+f_6)/2$ \\ \hline

$24$ & $f_8$    & $= f_7 \quad ;  \quad f_4 = f_3$ \\
    & $f_{11}$ & $= (f_1-f_2)/2+f_{12}$ \\
    & $f_{10}$ & $= (f_5-f_6)/2+f_9$ \\
    & $f_{13}$ & $= (-f_5+f_6-f_1+f_2)/2+f_{14}$ \\ \hline

$25$ & $f_7$     & $= f_8 \quad ;  \quad f_5 = f_6$ \\
    & $f_{11}$ & $= (f_1-f_2+f_3-f_4)/2+f_{12}$ \\
    & $f_9$    & $= (-f_3+f_4)/2+f_{10}$ \\
    & $f_{13}$ & $= (-f_1+f_2)/2+f_{14}$ \\ \hline

$26$ & $f_7$    & $= f_8 \quad ;  \quad f_5 = f_6$ \\
    & $f_{11}$ & $= (f_1-f_2+f_3-f_4)/2+f_{12}$ \\
    & $f_9$    & $= (-f_3+f_4)/2+f_{10}$ \\
    & $f_{13}$ & $= (-f_1+f_2)/2+f_{14}$ \\ \hline

$27$ & $f_6$    & $= f_5 \quad ;  \quad f_8 = f_7$ \\
    & $f_{10}$ & $= (f_3-f_4)/2+f_9$ \\
    & $f_{14}$ & $= (f_1-f_2)/2+f_{13}$ \\
    & $f_{12}$ & $= (-f_3+f_4-f_1+f_2)/2+f_{11}$ \\ 
\end{supertabular}
\end{center}

\end{appendix}

\end{document}